\documentclass[sigconf, screen]{acmart}

\usepackage[utf8]{inputenc}
\usepackage[T1]{fontenc}

\usepackage{graphicx}
\usepackage{subcaption}
\usepackage[export]{adjustbox}
\usepackage{textcomp}
\usepackage[hang,flushmargin]{footmisc}
\usepackage{url}
\usepackage{import}
\usepackage{color}

\usepackage{soul}
\usepackage{pifont}
\usepackage{enumitem}
\usepackage{multirow}
\usepackage{blindtext}
\usepackage{dirtytalk} 
\usepackage{tabularx} 
\usepackage{multirow}
\usepackage{listings} 
\usepackage{todonotes}
\usepackage{mathtools}
\usepackage{cleveref}
\usepackage{rotating}

\newcounter{svcounter}
\newcommand{\sv}[1]{%
    \stepcounter{svcounter}%
    {\color{red}\{SV\arabic{svcounter}: #1\}}%
}

\graphicspath{ {./Figures/} }
\usepackage[autostyle=true]{csquotes}
\MakeOuterQuote{"}

\lstdefinestyle{python}{ 
    xleftmargin=6.0ex,
    xrightmargin=2.0ex,
    numbers=left,
    frame=single
}
\definecolor{gray10}{gray}{.9}
\definecolor{arsenic}{rgb}{0.23, 0.27, 0.29}
\usepackage{color}

\RequirePackage{fontawesome}

\definecolor{gray50}{gray}{.5}
\definecolor{gray40}{gray}{.6}
\definecolor{gray30}{gray}{.7}
\definecolor{gray20}{gray}{.8}
\definecolor{gray10}{gray}{.9}
\definecolor{gray05}{gray}{.95}

\newlength\Linewidth
\def\findlength{\setlength\Linewidth\linewidth
    \addtolength\Linewidth{-4\fboxrule}
    \addtolength\Linewidth{-3\fboxsep}
}

\newenvironment{rqbox}{\par\begingroup
	\setlength{\fboxsep}{5pt}\findlength
	\setbox0=\vbox\bgroup\noindent
	\hsize=0.95\linewidth
	\begin{minipage}{0.95\linewidth}\normalsize}
	{\end{minipage}\egroup
	\textcolor{gray20}{\fboxsep1.5pt\fbox
		{\fboxsep5pt\colorbox{gray05}{\normalcolor\box0}}}
	\endgroup\par\noindent
	\normalcolor\ignorespacesafterend}

\begin{document}

\title{Does Microservices Adoption Impact the Development Velocity? A Cohort Study
}
\subtitle{(Registered Report)}

\author{Nyyti Saarim{\"a}ki}
   \affiliation{
   \institution{Tampere University\\} 
   \city{Tampere}
   \country{Finland}
 }
 \email{nyyti.saarimaki@tuni.fi}

\author{Mikel Robredo}
   \affiliation{
   \institution{University of Oulu\\} 
   \city{Oulu}
   \country{Finland}
 }
 \email{mikel.robredomanero@oulu.fi}

 \author{Sira Vegas}
 \affiliation{
   \institution{Universidad Politécnica de Madrid\\} 
   \city{Madrid}
   \country{Spain}
 }
 \email{svegas@fi.upm.es}

 \author{Natalia Juristo}
 \affiliation{
   \institution{Universidad Politécnica de Madrid\\} 
   \city{Madrid}
   \country{Spain}
 }
 \email{natalia@fi.upm.es}

 \author{Davide Taibi}
 \affiliation{
   \institution{University of Oulu\\} 
   \city{Oulu}
   \country{Finland}
 }
 \email{davide.taibi@oulu.fi}

 \author{Valentina Lenarduzzi}
 \affiliation{
   \institution{University of Oulu\\} 
   \city{Oulu}
   \country{Finland}
 }
 \email{valentina.lenarduzzi@oulu.fi}

\renewcommand{\shortauthors}{Saarim\"{a}ki, et al.}



\begin{abstract}

[Context] Microservices enable the decomposition of applications into small and independent services connected together. 
The independence between services could positively affect the development velocity of a project, which is considered an important metric measuring the time taken to implement features and fix bugs. However, no studies have investigated the connection between microservices and development velocity.

[Objective and Method] The goal of this study plan is to investigate the effect microservices have on development velocity. The study compares GitHub projects adopting microservices from the beginning  and similar projects using  monolithic architectures. We designed this study using a  cohort study method, to enable obtaining a high level of evidence.

[Results] The result of this work enables the confirmation of the effective improvement of the development velocity of microservices. Moreover, this study will contribute to the body of knowledge of empirical methods being among the first works adopting the cohort study methodology.  

\end{abstract}


\maketitle

\section{Introduction}

Microservices have become popular over the last years as they are considered to have several benefits over the traditional monolithic structure. They divide a project into smaller autonomous services each having a clear purpose~\cite{fowler2014microservices}. Therefore, microservice architecture has a significant effect on the overall development of a project and it allows independent and concurrent development and deployment~\cite{Waseem2020}, reusability~\cite{fowler2014microservices}, and enhanced scalability~\cite{Dragoni2018}. In theory, this should make also the development process faster. Development velocity is an important measure in current software projects and a higher velocity implies faster implementation of features and quicker resolution of issues. However, there are no studies investigating the connection of microservices and development velocity.

The goal of this study is to investigate the effect microservices have on development velocity.  We hypothesize that microservice projects are faster to develop than monolithic ones. To investigate this, we have planned a cohort study comparing actively developed projects adopting microservices from the beginning and similar projects using a monolithic structure. 

The proposed study plan covers all aspects required for conducting the study, including data sources, variable definitions, and data analysis. The study is designed as a cohort study because of the causal nature of the goal. As a result, the outcomes of the study are expected to provide high-level evidence on the topic.

\textbf{Paper structure:} Section~\ref{sec:background} describes the background and the related work. Section~\ref{sec:EmpiricalStudy} describes our study design and the execution. Section~\ref{sec:ThreatsValidity} discusses the possible threats. Section~\ref{sec:Conclusion} draws conclusions and future work.


\section{Background and Related Work}
\label{sec:background}



\subsection{Microservices}
The Microservice style is derived from Service-Oriented Architecture (SOA), where services have dedicated responsibilities but are not independent~\cite{fowler2014microservices, dragoni2017microservices}. In SOA, the individual services are neither full-stack (e.g., the same database is shared among multiple services) nor fully autonomous (e.g., service A depends on service B). On the contrary, microservices systems are decentralized systems composed of a large number of small independent services that communicate through different lightweight mechanisms~\cite{Pahl2018}. 

Microservices are relatively small and autonomous services deployed independently, with a single and clearly defined purpose~\cite{fowler2014microservices}. Microservices enable vertically decomposing applications into a subset of business-driven independent services. Each service can be developed, deployed, and tested independently by different development teams using different technology stacks. Microservices have a variety of advantages \cite{jamshidi2018microservices}. They can be developed in different programming languages, can scale independently from other services, and can be deployed on the hardware that best suits their needs. Moreover, because of their size, they are easier to maintain and more fault-tolerant since the failure of one service will not disrupt the whole system, which could happen in a monolithic system~\cite{fowler2014microservices}.


\subsection{Docker}

Docker\footnote{Docker: \url{https://www.docker.com/}} is a platform for developing, shipping and running applications in loosely isolated environments called containers. A Docker container includes everything needed to run an application and, therefore, it enables the developer to develop on their platform on choice without having to worry about where the program is deployed~\cite{jangla2018accelerating}. This is one of the several reasons why Docker is a popular tool among developers, and it was voted as the most important tool in StackOverflow's developer survey in 2022\footnote{\url{https://survey.stackoverflow.co/2022/\#most-popular-technologies-tools-tech-prof}}. The tool is commercial but is has a free version for open-source communities and individual developers.

A dockerized project can consist of one or several containers. The containers can communicate with each other but the implementation of each container is otherwise independent of other containers. Therefore, Docker is one way of creating microservices. In practice, each container has a file called \verb|Dockerfile|. It is a ''is text document that contains all the commands a user could call on the command line to assemble an image.''~\cite{docker2023}.

\subsection{Development velocity}

Development velocity estimates the amount of productive work a developer team can complete in a given time frame. 

The influence of microservices on velocity can vary depending on different factors. Here are some ways in which the adoption of microservices can affect velocity:

\textit{Independent Development and Deployment:} Microservices facilitate the autonomous development and deployment of individual services. This enables teams to concurrently work on different services, reducing dependencies and bottlenecks. Consequently, development cycles can be expedited, allowing for quick iteration and delivery of new features or improvements~\cite{Soldani2018}\cite{fowler2014microservices}.

\textit{Enhanced Scalability: }Microservices architecture empowers the independent scaling of individual services based on their specific requirements. This flexibility enables teams to optimize performance and responsiveness, ensuring efficient utilization of resources. Fine-grained scaling aligned with demand can bolster velocity by effectively handling the increased workload~\cite{Dragoni2018}.

\textit{Concurrent Development and Testing:} Microservices architecture facilitates parallel development and testing as services can be developed and tested in isolation. Teams can independently work on different services, enabling concurrent progress. This significantly speeds up the development lifecycle, as changes and updates can be implemented in parallel, reducing overall development time~\cite{Waseem2020}.

\textit{Reusability and Modularity:} Microservices encourage the creation of small, reusable components that can be shared across services. This promotes reusability and modularity, accelerating development by leveraging existing services, libraries, and frameworks. Developers can build upon existing functionalities, reducing redundant efforts and hastening the development process~\cite{fowler2014microservices}.

Ultimately, the influence of microservices on velocity depends on the effectiveness of the architecture design, the maturity level of microservices adoption, the skills and experience of the development teams, and the efficiency of supporting processes and tools~\cite{Soldani2018}\cite{TaibiIEEE2017}.

To the best of our knowledge, this is the first study empirically investigating the impact of microservices on velocity.

\section{The Empirical Study: Design and Execution Plan}
\label{sec:EmpiricalStudy}
In this Section, we describe our empirical study focusing on the design and the execution plan.

\subsection{Goal and Research Questions}

The goal of the study is to understand the effect of Docker-based microservice architecture on the development velocity during the early stages of the evolution of a software project. Our goal is answered by the following research question:

\begin{center}	
	\begin{rqbox}
		\textbf{RQ.} \emph{Do projects adopting microservices from the beginning have higher development velocity than monolithic projects?}
  \end{rqbox}	 
\end{center}

The hypothesis is that using Docker-based microservice architecture may help to have a higher development velocity. 

Decomposing software  projects into independent dockerized microservices are expected to  accelerate the development process as each service can be independently deployed, and  optimized  according to their needs~\cite{VelicityBook}. Despite Docker being a popular tool among developers, there are no studies on the topic exploring the tool's effect on velocity. Therefore, this is the first study that empirically investigates the impact of microservices on velocity. 



\subsection{Study Design}

\begin{figure}[h!]
    \centering
    \includegraphics[width=\columnwidth]{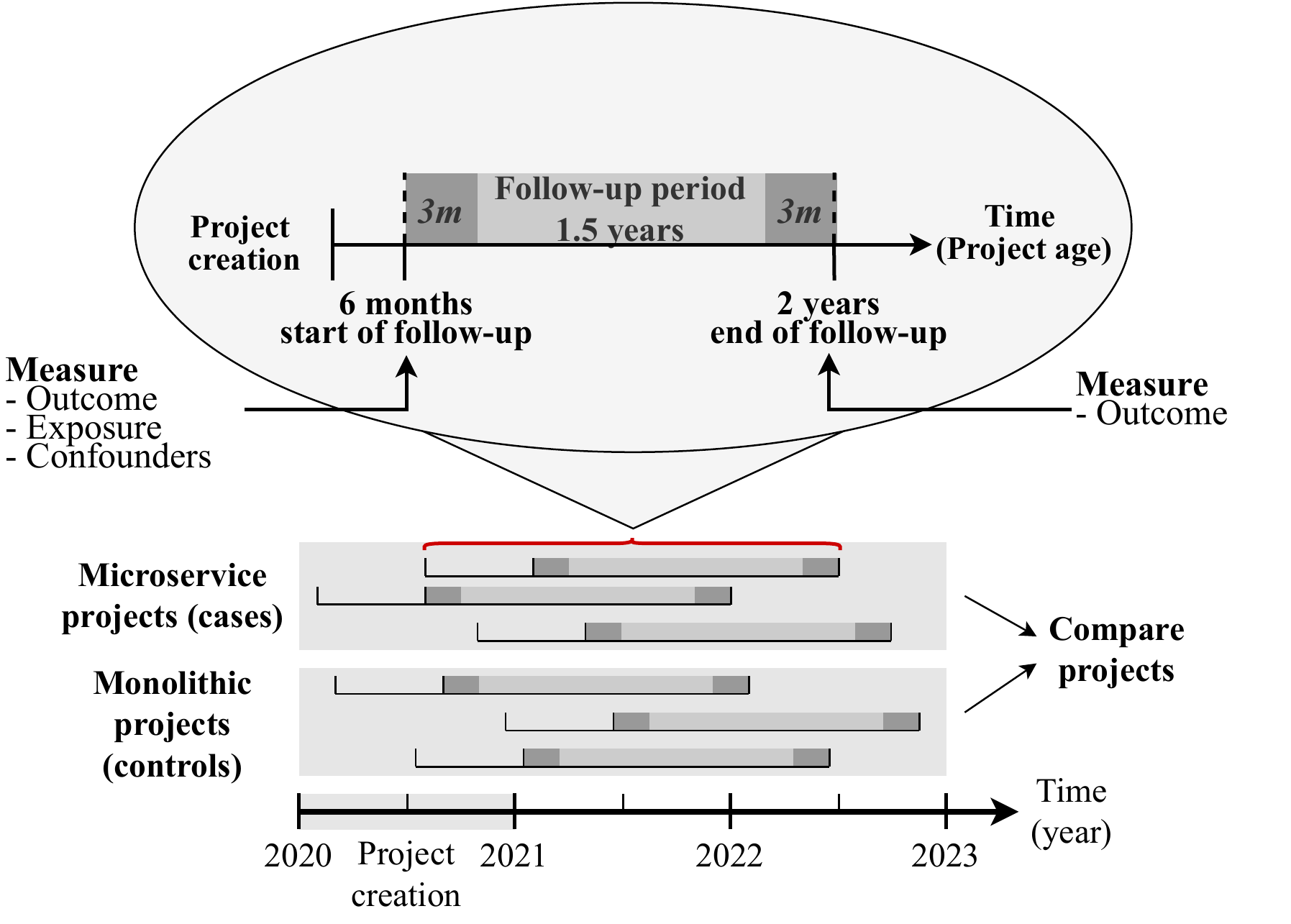}
    \caption{The design of the observational study.}
    \label{fig:study_design}
\end{figure}

The research question of this paper is causal in nature and the most suitable methodology for studying causality is a controlled experiment. However, the data is observational and historical which prevents us from conducting one. Thus, the study was designed as a \textit{retrospective cohort study} which is an analytical observational study methodology capable of obtaining high-level evidence from such data. The overall design of our study is presented in Figure~\ref{fig:study_design}.

A cohort study~\cite{grimes2002cohort, gordis} investigates whether an exposure (independent variable) causes an outcome (dependent variable) by comparing the outcomes of two or more groups of study subjects with different levels of exposure. It carefully selects a study population and follows it over a defined time period to see what naturally happens. The subjects are selected from the source population (source of data) using unambiguous eligibility criteria. They are needed to ensure the study population is capable to answer the research questions in the context of the population of interest. 

As the methodology is observational, the results can be influenced also by other factors than the exposure and outcome, such as project age or size. Therefore, in cohort studies, it is crucial to identify these factors from literature or using domain knowledge, measure them, and control for their effect. 

The exposure, confounders, and outcome are measured for all study subjects at the start of the follow-up. The outcome is measured at the end of the follow-up period. The gathered data is then used to analyze whether there is a relationship between the exposure and the outcome. Note that the outcome is measured also at the start of the study to control for the starting point. This provides a temporal framework that makes it possible to assess causality.



\subsection{Setting}

The study investigates open-source projects from GitHub which are created between 2020-2021. The cases of the study are projects which have adopted a  microservice architecture within six months of their creation while the controls of the study have a monolithic structure.

 The projects are tracked for 18 months starting from when they are half a year old until they reach two years of age. Therefore, the study subjects have different data measurement dates, but the same fixed length of follow-up time (Figure~\ref{fig:study_design}).


\subsection{Subjects}


The subjects of the study are open-source software projects. To ensure the study subjects can be used to answer the study goals, eligibility criteria are needed.

The \textbf{inclusion criteria} determines what is required from each subject in order to be considered as a part of the study. 

\begin{itemize}
    \item Open-source project created in GitHub between 2020-2021
    \item Uses GitHub to track issues
\end{itemize}


\textbf{Exclusion criteria} defines subjects which are excluded from the study at the start of the follow-up period. We applied the following criteria:

\begin{itemize}
    \item \textit{Project has one or two Dockerfiles}. Projects having only one or two Dockerfiles are purely neither controls nor cases. We consider projects having less than three to not have fully adopted the microservices, or not being large enough to be considered in this study. We do not consider files like dockerized databases or volumes in this count. 
    
\end{itemize}

In addition to selection criteria, we define a \textbf{loss to follow-up criteria}. The criteria define the subjects which are excluded from the study based on their activity during the follow-up period. Subjects not meeting the criteria at the end of the follow-up period are considered drop-outs and excluded from the final study subjects.

\begin{itemize}
    
    \item \textit{The overall monthly trend of commits is decreasing during the follow-up}. Ensures the project is developed during the follow-up which is required to be able to detect the outcome. Inspecting the trend includes projects with different resources in the data set.
    
    \item \textit{The overall monthly trend of issues is decreasing during the follow-up}. Ensures tracking issues using a well-integrated part of the development process.
    
    \item \textit{Use of Docker is interrupted during the follow-up period.} 
    Projects which at any point before the end of the follow-up period introduced Docker and then removed it. This ensures the projects maintain the exposure through the follow-up period and the projects are free of any potential effects of previous Docker usage. 
    
\end{itemize}



\subsection{Variables}

\begin{figure}
    \centering
    \includegraphics[width=\columnwidth]{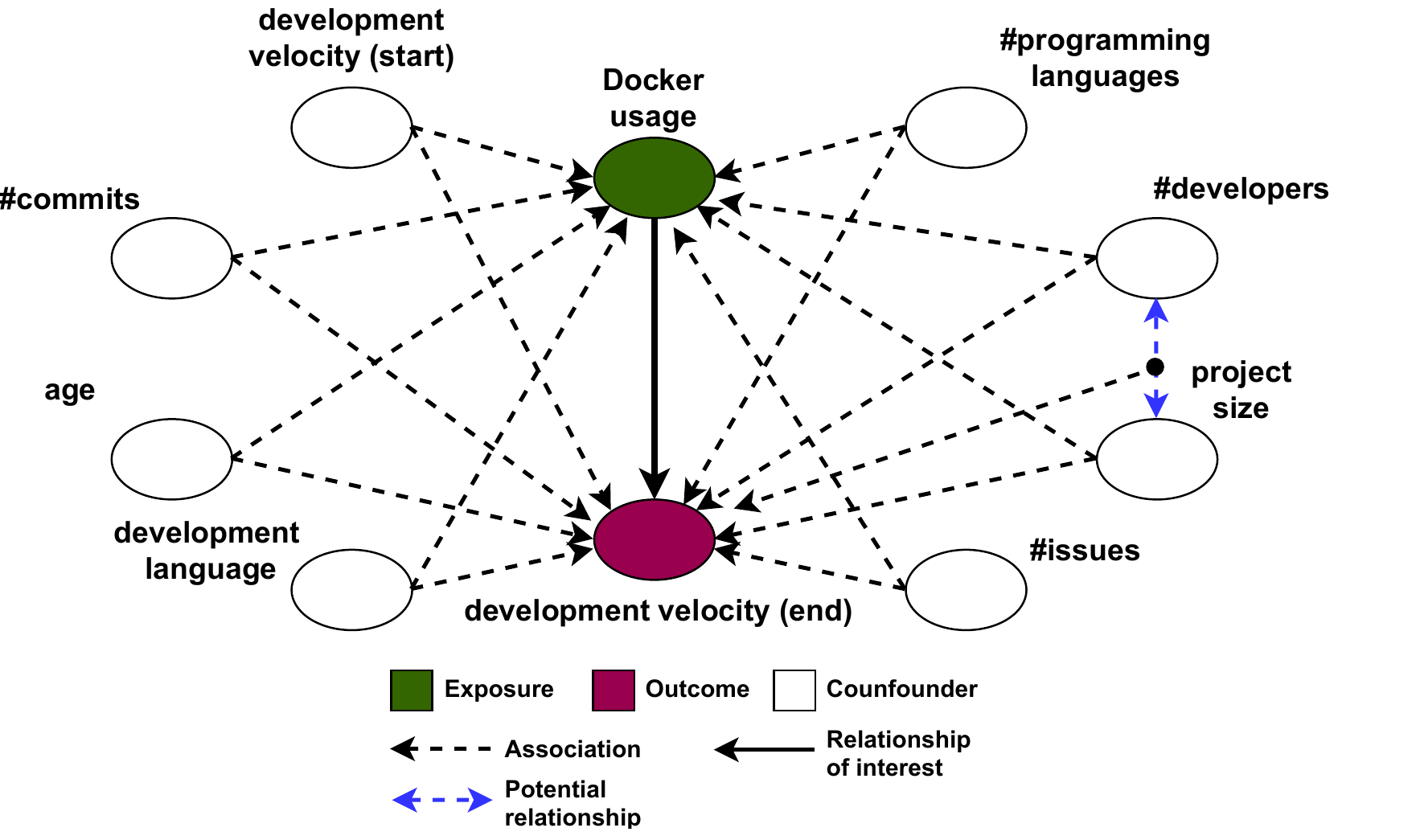}
    \caption{Graph visualizing the included variables and their (hypothesized) connections.}
    \label{fig:DAG}
\end{figure}

The variables included in the study are described below. The hypothesized relationships between the variables which are included in the study are visualized in Figure~\ref{fig:DAG}.

The \textbf{independent variable} is \textit{adoption of microservice architecture}. It is a boolean variable indicating if a project uses the microservice architecture or not. We focus on projects implementing the microservices using Docker and define a project to have adopted microservice architecture if the repository contains at least three docker files at the age of six months. Three docker files are considered to be the minimum threshold for a system to be considered distributed.

\begin{figure}
    \centering
    \includegraphics[width=0.7\columnwidth]{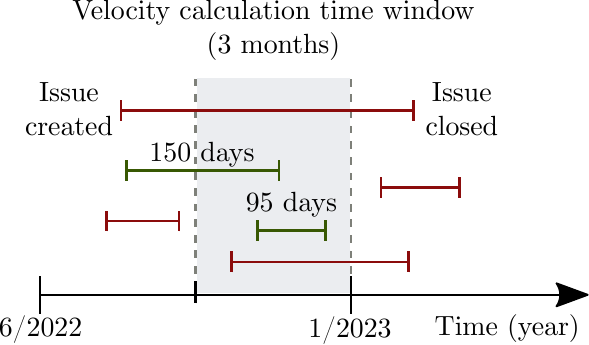}
    \caption{Velocity calculation. The issues closed in the selected time window (green) are included in the velocity calculation. The velocity in the figure is (150 + 95) days / 2 = 122.5 days.}
    \label{fig:velocity_calculation}
\end{figure}

We consider the \textbf{dependent variable} as the \textit{development velocity of a project} at the end of the follow-up period. The velocity of a project is the average time taken to close issues within a three month period. Therefore, lower velocity indicates faster issue fixing time. Figure~\ref{fig:velocity_calculation} presents an example of the calculation while the measurement period is visualized in Figure~\ref{fig:study_design}.

Based on our domain knowledge, the following preliminary set of \textbf{confounding variables} is considered for the study. Even though collecting the planned variables should be possible, the set of considered confounders might be altered during data collection. All confounding variables are measured at the start of the follow-up period.

 \begin{itemize}[leftmargin=*]
    \item \textit{Start velocity}. The velocity of the project before the follow-up period. 
    
    \item \textit{Size of the project}: The number of lines of code in the repository. 
    
    \item \textit{Development language}: The main development language reported by GitHub. Different languages might have a different impact on developers' productivity, and therefore on velocity. High-level languages might enable faster development (i.e. higher velocity) while lower-level languages might require a longer time between deployments (e.g. C or C++).
    
    \item \textit{Number of programming languages}: a higher number of different programming languages might fragment the development community, and therefore reduce velocity in case no developers are able anymore to modify a service written in a specific language. 
    
    \item \textit{Age}: The number of days from the first commit to the start of the follow-up.

    \item \textit{Number of commits}: Number of commits to the main branch of the repository by the start of the follow-up reported by GitHub. Each commit might create or fix an issue, and therefore, it could affect the velocity. 

    \item \textit{Number of issues}: The number of created issues (open and closed) at the start of the follow-up. The number of issues reported for the

    \item \textit{Number of developers}: The number of persons contributing to the project according to GitHub. In theory, more developers should mean higher development velocity.
projects affect its velocity.
    
\end{itemize}

In addition to confounders, we have identified the following potential relationships between them which have potential effects on the outcome variable. These relationships will not be included in the study as additional variables. 
\begin{itemize}
    \item \textit{Size $/$  \#Developers}: A proxy for the amount of code a single developer produces. 
\end{itemize}

\subsection{Data sources and measurement}
\label{sec:context}



The source population (list of potential projects) will be extracted from the World of Code (WoC) version U (updated 10/2021) or the following ones, based on their availability when conducting the study. It is a Free/Libre Open Source Software (FLOSS) ecosystem and a computational and statistical infrastructure that aims to provide an operational, research-ready, updatable, and expandable dataset~\cite{mockus2023woc}. The huge dataset is curated by fully collecting and cross-referencing project objects (e.g., authors, projects, commits, and blobs) mainly from public version control systems. WoC enables getting information on project commits, blogs, and files. 

The value of the independent variable (microservice adoption) is determined using data from version control history. We crawl the commit history of the repositories to determine the dates on which dockerfiles were added or modified. From that information, we are able to determine the number of dockerfiles in the project at any given date.

The data for determining the development velocity (dependent variable) is gathered from GitHub using its REST API. We will crawl information about all issues of the included repositories. Specifically, we will gather the creation, update, and closing date for each issue.

The confounding variables are collected also from the projects' GitHub repositories. As GitHub API does not provide the size of the projects, we will clone all included projects. If needed, we will analyze the repository using a tool to determine the value of a confounder.


\subsection{Study size}
\label{sec:sample_size}

Being able to detect a difference between two groups requires a sufficient number of subjects. Despite the data being mined from GitHub, which has a large set of projects, the strict eligibility criteria of the study might result in a small set of suitable projects. To understand and mitigate this risk, we adopt power analysis which is a statistical calculation conducted to determine the minimum required sample size for detecting significant effects from the data. It ensures the study results are valid with a given level of confidence and can detect meaningful effects.

Its calculation requires defining the power, significance level, and estimated effect size for the study. Power is the probability of rejecting the null-hypothesis when it is false while significance level determines the probability of rejecting the null-hypothesis when it is true. We used value 80\% for the power and 5\% for the significance level, both commonly adopted values. The size of the relevant effect in this study was determined as small (Cohen's d $\ge$ 0.2~\cite{cohen2013statistical}). Even Docker is adopted mainly for other reasons than enhancing development velocity., it is expected to have an effect also on the velocity. Therefore, we considered negligible changes irrelevant to this study. 

Due to the strict eligibility criteria, the number of cases in the obtained data set could be lower than expected for a 1:1 case-to-control ratio. Having up to four controls for each case can increase the statistical power of a study~\cite{song2010observational}. Therefore, we calculated the required sample sizes for several case-to-control ratios. The analysis was conducted using SPSS's (version 29) power analysis for ''Independent-samples T Test'' functionality and t results from the analysis are presented in Table~\ref{tab:power_analysis}.

\begin{table}[h]
    \centering
    \begin{tabular}{l|cccc}
     & \multicolumn{4}{c}{\textbf{Case to control ratio}} \\
     & \textbf{1:1} & \textbf{1:2} & \textbf{1:3} & \textbf{1:4}  \\
    \hline
    \textbf{\#Cases} & 394 & 295 & 263 & 246 \\
    \textbf{\#Controls} & 394 & 590 & 787 & 983 \\
    \hline
    \textbf{Total} & 788 & 885 & 1,050 & 1,229 \\
    \hline
    \multicolumn{5}{l}{Power=0.8, p=0.05, Effect size=0.2}
    \end{tabular}
    \caption{Results from the power calculation for the different case-to-control ratios.}
    \label{tab:power_analysis}
\end{table}

As the potential controls for this study are projects in GitHub using a monolithic structure, the number of potential controls is expected to be more than quadruple compared to the potential cases. In such case, we will consider performing random sampling for the selection of the control projects. 



\subsection{Statistical analysis}
\label{sec:data_analysis}

The analysis will be conducted using SPSS (version 29) and R (version 4.1.1).

\begin{figure}
    \centering
    \includegraphics[width=0.6\columnwidth]{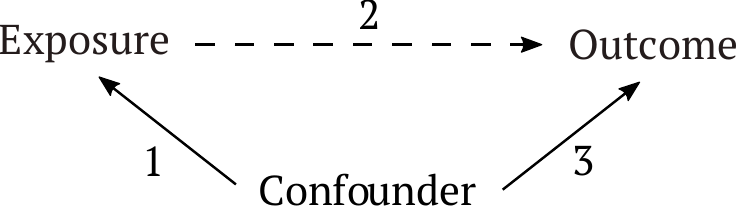}
    \caption{The relationships between exposure, outcome, and confounder.}
    \label{fig:variable_relationships}
\end{figure}

The connections between exposure, outcome, and confounder are visualized in Figure~\ref{fig:variable_relationships}. The data analysis of the study investigates several aspects of the figure.

\subsubsection{Crude analysis}

First, we determine the unadjusted relationship between the cases and controls (arrow 2). The analysis does not include any external variables and it compares the velocity between cases and controls. The results serve as base knowledge and justification for further analysis including the external variables. The groups are compared using independent samples t-test and Cohen's d effect size which both require (approximate) normality and homogeneity of variances. If these requirements are not fulfilled, data transformation is performed in order to fulfill them. In case the data transformation does not provide approximated normality, the comparison is conducted using the non-parametric Mann-Whitney test and Cliff's delta effect size. 

\subsubsection{Treating confounders}

We investigate if the included variables are actually confounders. A confounder needs to have a relationship to both exposure and outcome. First, we analyze if there is a relationship between the confounders and the exposure (arrow 1) by comparing the values of potential confounders between the case and control groups. 

The treatment method for the included confounders is chosen individually. The choices are made based on the characteristics of the collected data. Below we present the common techniques for treating confounding~\cite{grimes2002bias}, however, as the methods cannot be used in all circumstances, the used techniques depend on the observed scenario.

\begin{itemize}[leftmargin=*]
    \item \textit{Restriction} removes the effect of a confounding variable by ensuring all subjects are exposed to the same level of confounding. It is the simplest of the methods as in practice this is done by adding exclusion criteria and no further analysis is needed. However, after restricting a variable, its effect cannot be assessed.

    \item \textit{Matching} ensures the study population contains $n$ similar controls for each case and is done in the design phase of the study. This ensures the groups are exposed to the effect confounders similarly. The matching criteria can consist of one or a combination of several variables and the matching can be conducted using several different algorithms~\cite{RosenbaumPaulR2010DoOS}. However, matching prevents studying the variables used in the process.

    \item \textit{Stratification} ensures in the analysis phase of the study that subjects with different levels of confounder have similar effects between the cases and controls. The study subjects are divided into groups (or strata) based on the level of confounding they are exposed to. The analysis is conducted separately for each group and if the results differ, an adjusted value is calculated, for example, using meta-analysis~\cite{borenstein2021introduction}.

    \item \textit{Statistical adjustment} uses mathematical models to determine the relationship on interest while controlling for the effect confounders. The confounders are added to the considered models as additional independent covariates. To determine the covariates' impact on the results for the outcome variable, we will consider potential methodologies of regression analysis based on the distributional assumptions derived from the data. Similarly, based on the assumed distribution, we will analyze the variance of the dependent variable subject to the independent variable while accounting for the rest of covariates.
     
    To assess the significance of the considered models, we utilize existing information criteria methods such as \textit{Akaike Information Criterion} (AIC) or \textit{Bayesian Information Criterion}. Moreover, we consider performing \textit{Backward Selection} (BS) procedure based on the results from the mentioned information criteria to achieve the model that better describes the relationship between the dependent and independent variables.

\end{itemize}





\section{Threats to Validity}
\label{sec:ThreatsValidity}

\textbf{Construct Validity.} The measurement of microservice adoption is a threat in the study as version control does not provide data for validating its actual usage. Thus, we will rely on the number of dockerfiles in the repository. We try to mitigate this threat by requiring at least three dockerfiles and excluding volumes etc from the count. However, this does not guarantee active microservice adoption.

\textbf{Internal Validity.} The projects are collected from a version control system that automatically collects and logs the data. This ensures the similarity of the data collection between cases and controls. Additionally, GitHub has a large number of projects and we should be able to gather a data set that matches the minimum sample size determined in Section~\ref{sec:sample_size}. However, the exclusion and loss to follow-up criteria we set for this study might lead to a sharp reduction of analyzable projects. To mitigate this,  we would consider collecting projects from additional data sources, such as GitLab, as well as Jira or Bugzilla for issue tracking.

Cohort studies are sensitive to case and control groups not being comparable. We plan to ensure their similarity by suitable eligibility criteria and controlling for confounders using suitable methods such as restriction or matching.

The projects are followed from their creation until they are two years old and, therefore, they are in the same phase of their evolution. This could make the data collection process difficult as data measurement dates vary between the subjects. However, this is considered only a minor threat as GitHub automatically records the data.

The velocity of the project might differ between projects for several reasons. This is addressed by including external variables we have identified as potentially having an effect on it. To further mitigate this, we measure the development velocity also at the start of the follow-up and include the value in the analysis. The velocity of the project is calculated using a three month time window which might be inadequate for some projects, especially if the window is during holidays. The projects can also use some other issue tracking system than GitHub. However, we mitigate this threat by requiring a constant or growing trend of commits and issues during the follow-up. 

\textbf{External Validity.} The included projects are open-source projects from GitHub which reach a certain level of maturity within two years of their creation. Therefore, the results are generalizable to young and active open-source projects, that is, this study is not applicable to older projects with greater maturity. However, the real generalizability of the results will depend on the collected data.

\textbf{Conclusion Validity.} The planned data analysis follows the structure generally adopted in cohort studies. As we do not have the actual data yet, we do not know what are the most suitable data analysis methods. However, we have presented the general directions and alternatives for the planned analysis.
\section{Conclusion}
\label{sec:Conclusion}
The objective of this study plan is to examine the impact of microservices on development velocity. The study will involve a comparison between GitHub projects that initially implemented microservices and similar projects that utilized monolithic architectures. We have structured this study using a cohort study methodology to ensure a robust level of evidence.

The outcome of this research will validate the potential enhancement in development velocity achieved through the utilization of microservices. Additionally, this study will make a valuable contribution to the existing body of knowledge on empirical methods, as it will be one of the pioneering works adopting the cohort study methodology.

\bibliographystyle{ACM-Reference-Format}

\balance
\bibliography{sample-bibliography.bib}

\end{document}